\documentclass[twocolumn,emulatedapj]{aastex63}

\submitjournal{ApJ on 2019 November 15th}

\shorttitle{The X-ray outburst of the Galactic Center magnetar: SGR\,J1745$-$2900}
\shortauthors{Nanda Rea et al.}

\def\ltsima{$\; \buildrel < \over \sim \;$}
\def\lsim{\lower.5ex\hbox{\ltsima}}
\def\gtsima{$\; \buildrel > \over \sim \;$}
\def\gsim{\lower.5ex\hbox{\gtsima}}

\newcommand{\be}{\begin{equation}}
\newcommand{\en}{\end{equation}}

\def\nh{\hbox{$N_{\rm H}$}}
\def\flux {\mbox{erg cm$^{-2}$ s$^{-1}$}}
\def\lum {\mbox{erg s$^{-1}$}}

\def\arc{\mbox{$^{\prime\prime}$}}

\def\deg{\mbox{$^{\circ}$}}

\def\aa {1E\,1547$-$5408}

\def\xte{XTE\,J1810$-$197}

\def\galmag{SGR\,J1745$-$2900}
\def\sgras{Sgr\,A$^{\star}$}

\def\psr{PSR\,1622$-$4950}

\def\hbpsrtwo{PSR\,J1119$-$6127}

\newcommand\cxo{{\em Chandra}}
\newcommand{\R}{{\em ROSAT}}
\newcommand{\A}{{\em ASCA}}

\newcommand{\e}{{\em Einstein}}

\newcommand{\swift}{{\em Swift}}

\begin{document}

\title{The X-ray outburst of the Galactic Center magnetar over six years of \cxo\, observations}

\correspondingauthor{Nanda Rea}
\email{rea@ice.csic.es}

\author{N. Rea}
\affiliation{Institute of Space Sciences (ICE, CSIC), Campus UAB, Carrer de Can Magrans s/n, 08193, Barcelona, Spain}
\affiliation{Institut d'Estudis Espacials de Catalunya (IEEC), Carrer Gran Capit\`a 2--4, 08034 Barcelona, Spain}

\author{F. Coti Zelati}
\affiliation{Institute of Space Sciences (ICE, CSIC), Campus UAB, Carrer de Can Magrans s/n, 08193, Barcelona, Spain}
\affiliation{Institut d'Estudis Espacials de Catalunya (IEEC), Carrer Gran Capit\`a 2--4, 08034 Barcelona, Spain}

\author{D. Vigan\`o}
\affiliation{Institute of Space Sciences (ICE, CSIC), Campus UAB, Carrer de Can Magrans s/n, 08193, Barcelona, Spain}
\affiliation{Institut d'Estudis Espacials de Catalunya (IEEC), Carrer Gran Capit\`a 2--4, 08034 Barcelona, Spain}

\author{A. Papitto}
\affiliation{INAF--Osservatorio Astronomico di Roma, via Frascati 33, I-00076, Monteporzio Catone (RM), Italy}

\author{F. Baganoff}
\affiliation{Kavli Institute for Astrophysics and Space Research, Massachusetts Institute of Technology, Cambridge, MA 02139, USA}

\author{A. Borghese}
\affiliation{Institute of Space Sciences (ICE, CSIC), Campus UAB, Carrer de Can Magrans s/n, 08193, Barcelona, Spain}
\affiliation{Institut d'Estudis Espacials de Catalunya (IEEC), Carrer Gran Capit\`a 2--4, 08034 Barcelona, Spain}

\author{S. Campana}
\affiliation{INAF--Osservatorio Astronomico di Brera, via Bianchi 46, I-23807 Merate (LC), Italy}

\author{P. Esposito}
\affiliation{Scuola Universitaria Superiore IUSS Pavia, Palazzo del Broletto, piazza della Vittoria 15, 27100 Pavia, Italy}
\affiliation{INAF--Istituto di Astrofisica Spaziale e Fisica Cosmica, via E. Bassini 15, I-20133 Milano, Italy}

\author{D. Haggard}
\affiliation{Department of Physics, McGill University, 3600 University St., Montreal, QC H3A 2T8, Canada}
\affiliation{ McGill Space Institute, McGill University, Montreal, QC H3A 2A7, Canada}

\author{G. L. Israel}
\affiliation{INAF--Osservatorio Astronomico di Roma, via Frascati 33, I-00076, Monteporzio Catone (RM), Italy}

\author{S. Mereghetti}
\affiliation{INAF--Istituto di Astrofisica Spaziale e Fisica Cosmica, via E. Bassini 15, I-20133 Milano, Italy}

\author{R. P. Mignani}
\affiliation{INAF--Istituto di Astrofisica Spaziale e Fisica Cosmica, via E. Bassini 15, I-20133 Milano, Italy}
\affiliation{Janusz Gil Institute of Astronomy, University of Zielona G\'ora, ul. Szafrana 2, 65–516, Zielona G\'ra, Poland}

\author{R. Perna}
\affiliation{Department of Physics and Astronomy, Stony Brook University, Stony Brook, NY 11794, USA}

\author{J. A. Pons}
\affiliation{Departament de Fisica Aplicada, Universitat d' Alacant, Ap. Correus 99, E-03080 Alacant, Spain}

\author{G. Ponti}
\affiliation{INAF--Osservatorio Astronomico di Brera, via Bianchi 46, I-23807 Merate (LC), Italy}
\affiliation{Max Planck Institut f\"ur Extraterrestriche Physik, Giessenbachstrasse, D-85748 Garching, Germany}

\author{L. Stella}
\affiliation{INAF--Osservatorio Astronomico di Roma, via Frascati 33, I-00076, Monteporzio Catone (RM), Italy}

\author{D. F. Torres}
\affiliation{Institute of Space Sciences (ICE, CSIC), Campus UAB, Carrer de Can Magrans s/n, 08193, Barcelona, Spain}
\affiliation{Institut d'Estudis Espacials de Catalunya (IEEC), Carrer Gran Capit\`a 2--4, 08034 Barcelona, Spain}
\affiliation{Instituci\'o Catalana de Recerca i Estudis Avan\c{c}ats (ICREA), E-08010 Barcelona, Spain}

\author{R. Turolla}
\affiliation{Dipartimento di Fisica e Astronomia, Universit\'a di Padova, via F. Marzolo 8, I-35131 Padova, Italy}
\affiliation{Mullard Space Science Laboratory, University College London, Holmbury St. Mary, Dorking, Surrey RH5 6NT, UK}

\author{S. Zane}
\affiliation{Mullard Space Science Laboratory, University College London, Holmbury St. Mary, Dorking, Surrey RH5 6NT, UK}


\begin{abstract}

The magnetar \galmag, discovered at parsecs distance from the Milky Way central black hole, Sagittarius~A$^{\star}$, represents the closest pulsar to a supermassive black hole ever detected. Furthermore, its intriguing radio emission has been used to study the environment of the black hole, as well as to derive a precise position and proper motion for this object. The discovery of \galmag\, has opened interesting debates about the number, age and nature of pulsars expected in the Galactic center region. In this work, we present extensive X-ray monitoring of the outburst of \galmag\ using the \cxo\, X-ray Observatory, the only instrument with the spatial resolution to distinguish the magnetar from the supermassive black hole (2.4\arc\ angular distance). It was monitored from its outburst onset in April 2013 until August 2019, collecting more than fifty \cxo\ observations for a total of more than 2.3 Ms of data. Soon after the outburst onset, the magnetar emission settled onto a purely thermal emission state that cooled from a temperature of about 0.9 to 0.6 keV over 6 years. The pulsar timing properties showed at least two changes in the period derivative, increasing by a factor of about 4 during the outburst decay. We find that the long-term properties of this outburst challenge current models for the magnetar outbursts.

\end{abstract}



\section{Introduction} \label{sec:intro}


\begin{deluxetable*}{cccccccc}
\tablecaption{Log of the \cxo\ observations and spectral fitting results. 
} 
\tabletypesize{\scriptsize}
\tablenum{1}
\tablewidth{0pt}
\tablehead{
\colhead{Obs ID} 	     & \colhead{Start time (TT)} & 	\colhead{Exposure} & \colhead{Count rate} & \colhead{$kT_{{\rm BB}}$}			& \colhead{$R_{{\rm BB}}$}			& \colhead{Absorbed flux}	 & \colhead{Luminosity}	\\
	\colhead{}  & \colhead{(yyyy/mm/dd hh:mm:ss)} & \colhead{(ks)} &  		\colhead{(counts s$^{-1}$) }	 & \colhead{(keV)	}	   		& \colhead{(km)}			& \colhead{(10$^{-12}$~\flux)}	& \colhead{(10$^{35}$~\lum) }
	}
\startdata
14702                                       		&	2013/05/12 10:38:50  &  	13.7  & 	$0.545 \pm 0.006$	& 	$0.88\pm0.01$  & 	2.52$_{-0.08}^{+0.09}$	& 	16.3$_{-0.8}^{+1.0}$		& 	$4.9\pm 0.5$	\\ 
15040\tablenotemark{$_*$}		&      2013/05/25 11:38:37  &  	23.8  & 	$0.150 \pm 0.003$    & 	$0.85\pm0.02$  & 	$2.5\pm0.1$			& 	15.5$_{-1.3}^{+0.03}$	& 	$4.7\pm0.5$	\\
14703	     	                         		&	2013/06/04 08:45:16 &	16.8  &	$0.455 \pm 0.005$ 	&      $0.83\pm0.01$ 	 &    2.50$_{-0.08}^{+0.09}$      & 12.7$_{-0.6}^{+0.5}$		& 	$4.1\pm0.4$      \\ 
15651\tablenotemark{$_*$}		&    2013/06/05 21:32:38 	& 	13.8 	& 	$0.141 \pm 0.003$ 	 & 	$0.84\pm0.03$	 & 	$2.4\pm0.2$			& 12.5$_{-0.9}^{+0.07}$		& 	$3.8\pm0.4$	\\ 
15654\tablenotemark{$_*$}		&    2013/06/09 04:26:16 	&     9.0 	& 	$0.128 \pm 0.004$	 & 	$0.83\pm0.04$	 & 	$2.4\pm0.2$			 & 12.4$_{-0.9}^{+0.05}$		&	 $3.5\pm0.4$	\\ 
14946						&    2013/07/02 06:57:56	&    18.2	&	 $0.392 \pm 0.005$    & 	$0.85\pm0.01$ 	 & 	2.39$_{-0.08}^{+0.09}$	& 10.4$_{-0.7}^{+0.4}$		& 	$3.5\pm0.3$	\\
15041	  					&    2013/07/27 01:27:17	&    45.4	& 	$0.346 \pm 0.003$   	 & 	$0.824\pm0.008$  & 	2.16$_{-0.05}^{+0.06}$	& 9.2$_{-0.3}^{+0.2}$		& 3.0	$_{-0.4}^{+0.2}$   \\
15042	 					&    2013/08/11 22:57:58	&    45.7	& 	$0.317 \pm 0.003$     & 	$0.843\pm0.008$      & $2.09\pm0.05$		         & $8.2\pm0.3$				& 2.7$_{-0.4}^{+0.2}$    \\
14945						&    2013/08/31 10:12:46	&    18.2	& 	$0.290 \pm 0.004$     & 	$0.82\pm0.01$ 	& 1.89$_{-0.07}^{+0.08}$  	& 7.7$_{-0.4}^{+0.3}$		& $2.4\pm0.2$	\\
15043	        					&   2013/09/14 00:04:52	&    45.4	& 	$0.275 \pm 0.002$     & 	$0.812\pm0.008$ 	& 2.03$_{-0.05}^{+0.06}$ 		& 7.2$_{-0.3}^{+0.2}$		& 2.4	$_{-0.3}^{+0.2}$	\\
14944	        					&    2013/09/20 07:02:56	&    18.2	& 	$0.273 \pm 0.004$      & 	$0.81\pm0.01$ 	& 1.88$_{-0.07}^{+0.08}$		& $7.0\pm0.4$				& 2.3$_{-0.3}^{+0.2}$	\\
15044	        					&    2013/10/04 17:24:48	&    42.7	& 	$0.255 \pm 0.002$   	& 	$0.826\pm0.009$	& $1.98\pm0.06$ 		         & $6.4\pm0.2$				& 2.2$_{-0.3}^{+0.2}$	\\
14943					        &    2013/10/17 15:41:05	&    18.2	& 	$0.246 \pm 0.004$  	& 	$0.82\pm0.01$	& 1.95$_{-0.08}^{+0.09}$ 		& 6.1$_{-0.4}^{+0.2}$		& $2.1\pm0.3$		\\
14704	      				        &    2013/10/23 08:54:30	&    36.3	& 	$0.240 \pm 0.003$  	& 	$0.806\pm0.009$ 	& $1.94\pm0.06$			& 5.9$_{-0.3}^{+0.2}$		& $2.1\pm0.2$		\\
15045						&    2013/10/28 14:31:14	&    45.4	& 	$0.234 \pm 0.002$    & 	$0.817\pm0.009$	 & 	$1.83\pm0.05$	 	 	& 5.9$_{-0.2}^{+0.1}$	        & 2.0	$_{-0.2}^{+0.1}$	\\  
16508						&   2014/02/21 11:37:48  	&    43.4	& 	$0.156 \pm 0.002$      & 	$0.81\pm0.01$ 	 & 	$1.49\pm0.05$ 			& $3.7\pm0.1$			         & 1.3$_{-0.2}^{+0.1}$	\\  
16211	  					&   2014/03/14 10:18:27	&    41.8	& 	$0.149 \pm 0.002$      & 	$0.81\pm0.01$ 	 & 	$1.50\pm0.05$ 			& 3.4$_{-0.2}^{+0.1}$	         &  $1.2\pm0.2$		\\  
16212						&    2014/04/04 02:26:27	&    45.4	& 	$0.135 \pm 0.002$      & 	$0.81\pm0.01$ 	 & 	$1.38\pm0.05$ 			& 3.1$_{-0.2}^{+0.1}$		& $1.1\pm0.1$		\\  
16213						&    2014/04/28 02:45:05	&    45.0	& 	$0.128 \pm 0.002$      & 	$0.83\pm0.01$ 	& $1.37\pm0.05$ 			& $3.0\pm0.1$				& $1.0\pm0.1$		\\  
16214						&    2014/05/20 00:19:11	&    45.4	& 	$0.118 \pm 0.002$	  &    $0.81\pm0.01$ 	& $1.34\pm0.05$ 			& $2.7\pm0.4$				& $1.0\pm0.1$		\\  
16210	        				        &    2014/06/03 02:59:23	&    17.0	& 	$0.110 \pm 0.003$	  &    $0.84\pm0.02$ 	& 1.17$_{-0.06}^{+0.07}$		& 2.6$_{-0.3}^{+0.1}$		& $0.9\pm0.1$ 		\\  
16597						&    2014/07/04 20:48:12	&    16.5	& 	$0.097 \pm 0.002$      & 	$0.77\pm0.02$	& 1.36$_{-0.08}^{+0.09}$		& $2.1\pm0.4$				& $0.8\pm0.1$		\\
16215						&    2014/07/16 22:43:52 	&    41.5	& 	$0.090 \pm 0.001$	  & 	$0.81\pm0.01$ 	 & $1.16\pm0.05$ 			& $2.1\pm0.3$				& $0.73\pm0.08$		\\  
16216						&    2014/08/02 03:31:41 	&   42.7 	& 	$0.085 \pm 0.001$	  & 	$0.77\pm0.01$	& 1.27$_{-0.05}^{+0.06}$ 		& $1.9\pm0.2$				& $0.73\pm0.07$		\\  
16217	 					&    2014/08/30 04:50:12 	&    34.5 	& 	$0.079 \pm 0.002$	  & 	$0.77\pm0.01$		& $1.24\pm0.06$		& $1.8\pm0.2$				& $0.69\pm0.09$		\\  
16218						&    2014/10/20 08:22:28  	&    36.3	& 	$0.071 \pm 0.001$	  & 	$0.79\pm0.01$		& 1.09$_{-0.05}^{+0.06}$ 	& $1.7\pm0.2$				& $0.60\pm0.07$		\\  
16963						&    2015/02/13 00:42:04    &    22.7	& 	$0.056 \pm 0.002$ 	   & 	$0.79\pm0.02$		& 0.98$_{-0.06}^{+0.07}$	& $1.3\pm0.3$				& $0.46\pm0.06$		\\  
16966						&    2015/05/14 08:46:51    &    22.7	& 	$0.045  \pm 0.001$ 	  &   0.76$_{-0.02}^{+0.03}$ & 0.97$_{-0.08}^{+0.09}$ & $1.0\pm0.2$				& $0.40\pm0.05$		\\  
16965       					&    2015/08/17 10:35:47    &    22.7 	& 	$0.035  \pm 0.001$	  & 	$0.72\pm0.02$		& 0.92$_{-0.07}^{+0.09}$	& $0.7\pm0.2$				& $0.29\pm0.04$		\\  
16964						&    2015/10/21 06:04:57	&    22.6	& 	$0.026  \pm 0.001$	  & 	$0.74\pm0.03$		& 0.79$_{-0.08}^{+0.10}$ 	& $0.6\pm0.2$				& $0.24\pm0.03$		\\  
18055						&    2016/02/13 08:59:23	&    22.7	& 	$0.0133  \pm 0.0008$  &	 $0.71\pm0.04$		& 0.76$_{-0.11}^{+0.15}$	& $0.4\pm0.2$				& $0.18\pm0.03$		\\  
18056						&   2016/02/14 14:46:01	&    21.8	& 	$0.0146  \pm 0.0009$   &	0.75$_{-0.04}^{+0.05}$ & 0.68$_{-0.12}^{+0.15}$ & $0.4\pm0.2$				& $0.18\pm0.02$		\\  
18731						&   2016/07/12 18:23:59	&    78.4	& 	$0.0112  \pm 0.0004$   & 	$0.70\pm0.02$		& 0.70$_{-0.06}^{+0.07}$ 	& $0.31\pm0.02$			& $0.15\pm0.02$		\\  
18732						&   2016/07/18 12:01:38	&    76.6	& 	$0.0118  \pm 0.0004$   & 	$0.71\pm0.02$		& 0.72$_{-0.05}^{+0.06}$	& $0.35\pm0.02$			& $0.17\pm0.02$		\\  
18057						&   2016/10/08 19:07:12	&    22.7	& 	$0.0123 \pm 0.0008$    & 	$0.66\pm0.03$ 		& 0.79$_{-0.09}^{+0.12}$  & $0.26\pm0.02$			& $0.14\pm0.02$		\\  
18058						&   2016/10/14 10:47:43	&    22.7  	& 	$0.0122 \pm 0.0007$    & 	$0.64\pm0.03$ 		& 0.78$_{-0.10}^{+0.11}$  & $0.23\pm0.02$			& $0.13\pm0.02$		\\  
19726\tablenotemark{$_\dagger$}	 & 2017/04/06 03:46:05     &   28.2    &  	$0.0084 \pm 0.0003$   & 	$0.65\pm0.02$ 	& 0.66$_{-0.05}^{+0.06}$   &   0.17$_{-0.04}^{+0.01}$   	& $0.10\pm0.01$	\\  	
19727\tablenotemark{$_\dagger$}	 & 2017/04/07 04:56:10    &   27.8     &   -  &  - 	& - 	&  - 	& -  	\\  		 
20041\tablenotemark{$_\dagger$}     & 2017/04/11 03:50:13    &   30.9     &  -   &  - 	& - 	&  - 	  & -  \\  	
20040\tablenotemark{$_\dagger$}	 & 2017/04/12 05:17:13   &   27.5    	&  -   &  - 	& - 	&  - 	& -  	\\  	 
19703\tablenotemark{$_\dagger$}     & 2017/07/15 22:34:58   &   81.0    	&  	$0.0066 \pm 0.0002$    & 	$0.69\pm0.02$  & 0.57$_{-0.04}^{+0.05}$    &   0.19$_{-0.01}^{+0.004}$   	& $0.095\pm0.009$	 \\  	       
19704\tablenotemark{$_\dagger$}     & 2017/07/25 22:56:18   &  78.4 	&   -   &  - 	& - 	&  - 	& -  	\\  	 
20344\tablenotemark{$_\dagger$}	 & 2018/04/20 03:16:36   &  29.1 	&    	$0.00375 \pm 0.00018$  & 0.71$_{-0.03}^{+0.04}$ 	 & 0.40$_{-0.04}^{+0.05}$   &   0.11$_{-0.01}^{+0.003}$    & $0.051\pm0.005$	\\  	 
20345\tablenotemark{$_\dagger$}	 & 2018/04/22 03:30:07   &   28.5 	&   -  &  - 	& - 	&  - 	& -  	\\  	 
20346\tablenotemark{$_\dagger$}      & 2018/04/24 03:32:34  &    30.0	&   -  &  - 	& - 	&  - 	& -  	\\  
20347\tablenotemark{$_\dagger$}	 & 2018/04/25 03:36:14   &  32.7 	&   - &  - 	& - 	&  - 	& -  	\\  
21453\tablenotemark{$_\dagger$}     & 2019/03/29 04:02:30   &  30.0  	&  	$0.00256 \pm 0.00015$  & 0.69$_{-0.04}^{+0.05}$ 	& 0.36$_{-0.06}^{+0.08}$   &   0.070$_{-0.010}^{+0.010}$   & $0.036\pm0.004$	\\  	 
21454\tablenotemark{$_\dagger$}	 & 2019/03/30 05:33:34  & 30.5 	&   -  &  - 	& - 	&  - 	& -  	\\  
21455\tablenotemark{$_\dagger$}	 & 2019/03/31 05:19:02  &  30.0 	&   -  &  - 	& - 	&  - 	& -  	\\  	
21456\tablenotemark{$_\dagger$}	 &  2019/04/01 04:21:56  &   30.0  	&  -   &  - 	& - 	&  - 	& -  	\\  
22230\tablenotemark{$_\dagger$}	 & 2019/07/17 22:59:57  &   57.0  	& 	$0.00248 \pm 0.00013$   &  0.60$_{-0.03}^{+0.03}$ & 0.54$_{-0.08}^{+0.10}$ &  0.067$_{-0.060}^{+0.030}$   & $0.049\pm0.005$	\\  	
20446\tablenotemark{$_\dagger$}	 & 2019/07/21 00:08:32  &  57.6 	& -  &  - 	& - 	&  - 	& -  	\\  
20447\tablenotemark{$_\dagger$}	 &2019/07/26 01:40:35   &  57.6  	& -   &  - 	& - 	&  - 	& -  	\\  
20750\tablenotemark{$_\dagger$}	 & 2019/08/13 23:23:09  &   24.3  	& 	$0.00254 \pm 0.00020$   &  	0.62$_{-0.04}^{+0.04}$ 	& 0.49$_{-0.10}^{+0.15}$ 	 &  	 0.072$_{-0.010}^{+0.010}$      & 	$0.047\pm0.005$\\  	
22288\tablenotemark{$_\dagger$}	 & 2019/08/15 23:29:20  &   24.2  	& -  &  - 	& - 	&  - 	& -  	\\  
20751\tablenotemark{$_\dagger$}	 &2019/08/19 22:51:15  &    24.3  	& -   &  - 	&  - 	& - 	&  - 	\\
\hline
\hline
\enddata
\tablenotetext{*}{\cxo\ ACIS-S grating observations. $\dagger$ New unpublished observations. Fluxes and luminosities are in the 0.3--10\,keV energy range. Observations where a dash ($-$) is present, were merged with the above ones in the spectral modelling.}
\label{tab:spectralfits}
\label{tab:log}
\end{deluxetable*}

Due to the coupling of extreme gravitational fields with very strong magnetic fields, neutron stars are among the most interesting celestial objects. At the magnetic extreme of the pulsar population, some thirty sources were discovered in the past decades, collectively labelled as magnetars (see Kaspi \& Beloborodov 2017 for a recent review). These objects are typically characterized by: i) peculiar flaring/bursting activity on several timescales and luminosities ($L\sim10^{38}-10^{46}$\,\lum~ during $0.1-500$\,s), ii) long-term outburst activity, during which for months to years their persistent luminosity is enhanced by several orders of magnitudes (Coti Zelati et al. 2018), iii) relatively slow rotational periods compared to those of the isolated pulsar population (spin periods typically in the $0.3-10$\,s range), and iv) surface dipolar magnetic fields generally estimated to be of the order of $10^{13}-10^{15}$\,G. These properties lead to the idea of magnetars being powered by their large magnetic energy (Thompson \& Duncan 1995, 1996). Studies of  magnetar outbursts in the past decades (Perna \& Pons 2011; Pons \& Perna 2011; Pons \& Rea 2012; Vigano' et al. 2013; Gourgouliatos, \& Cumming 2014; Wood \&  Hollerbach 2015; Lander \& Gourgouliatos 2019) have led to a deeper understanding of the magnetar phenomenology, and in particular of the physics of the surface cooling after such a large energy injection (Pons \& Rea 2012; Li, Levin \& Beloborodov 2016). They also allowed the discovery of low-field magnetars (Rea et al. 2010, 2012a, 2013a, 2014), of magnetar-like emission in other neutron star classes such as Central Compact Objects (D'Ai et al. 2016; Rea et al. 2016) and canonical rotational powered pulsars (Gavriil et al. 2006; Kumar \& Safi-Harb 2006; Archibald et al. 2016). However, many questions still remain to be answered, such as the mechanism which triggers the outburst emission, the role of the magnetic field helicity inside the star and in the magnetosphere, and the effects of the outbursts on the long-term quiescent luminosity of these objects (Carrasco et al. 2019). The recent discovery of magnetar-like emission in sources not previously counted as magnetars, led to question the exact definition of a magnetar, as well as the birth properties and number of sources showing magnetar-related emission.\\
In this general context happened the discovery of a powerful magnetar as the closest known pulsar to the Milky Way central supermassive black hole, Sagittarius~A$^{\star}$ (\sgras).

\galmag\ was discovered on 2013 April 25, with the detection of a magnetar-like burst in the soft gamma-rays by the {\swift}-BAT instrument (Kennea et al. 2013a). Follow-up observations revealed a bright X-ray counterpart ($L_X \sim 5 \times 10^{35}$ erg~s$^{-1}$ for an assumed distance of 8.3\,kpc), with the striking feature of being located at an angular distance of only 2.4\arc\ from \sgras, resulting in a projected separation of 0.097\,pc from the central supermassive black hole (Rea et al. 2013b). Coherent pulsations at a spin period of $\sim$3.76\,s were detected both in the X-ray (Kennea et al. 2013b; Mori et al. 2013; Rea et al. 2013b; Kaspi et al. 2014) and in the radio band (Shannon \& Johnston 2013; Lynch et al. 2015; Pennucci et al. 2015), making  \galmag\ the fourth confirmed radio-loud magnetar alongside \xte, \aa\, and \psr\, (Camilo et al. 2006, 2007; Levin et al. 2010), with a dipolar surface magnetic field of $B\sim2\times10^{14}$\,Gauss.

\begin{figure*}
\includegraphics[width=9cm]{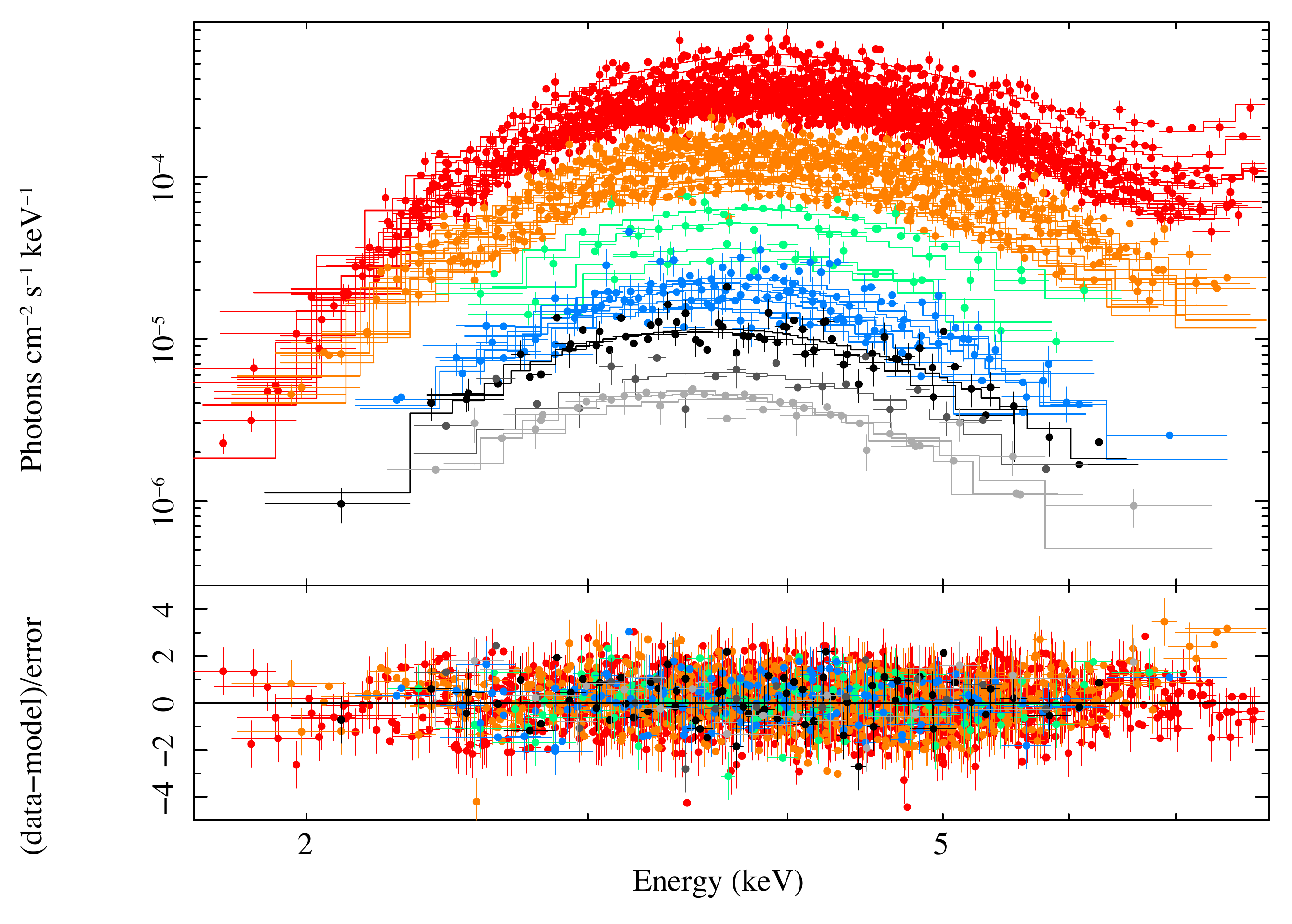}
\includegraphics[width=8.8cm]{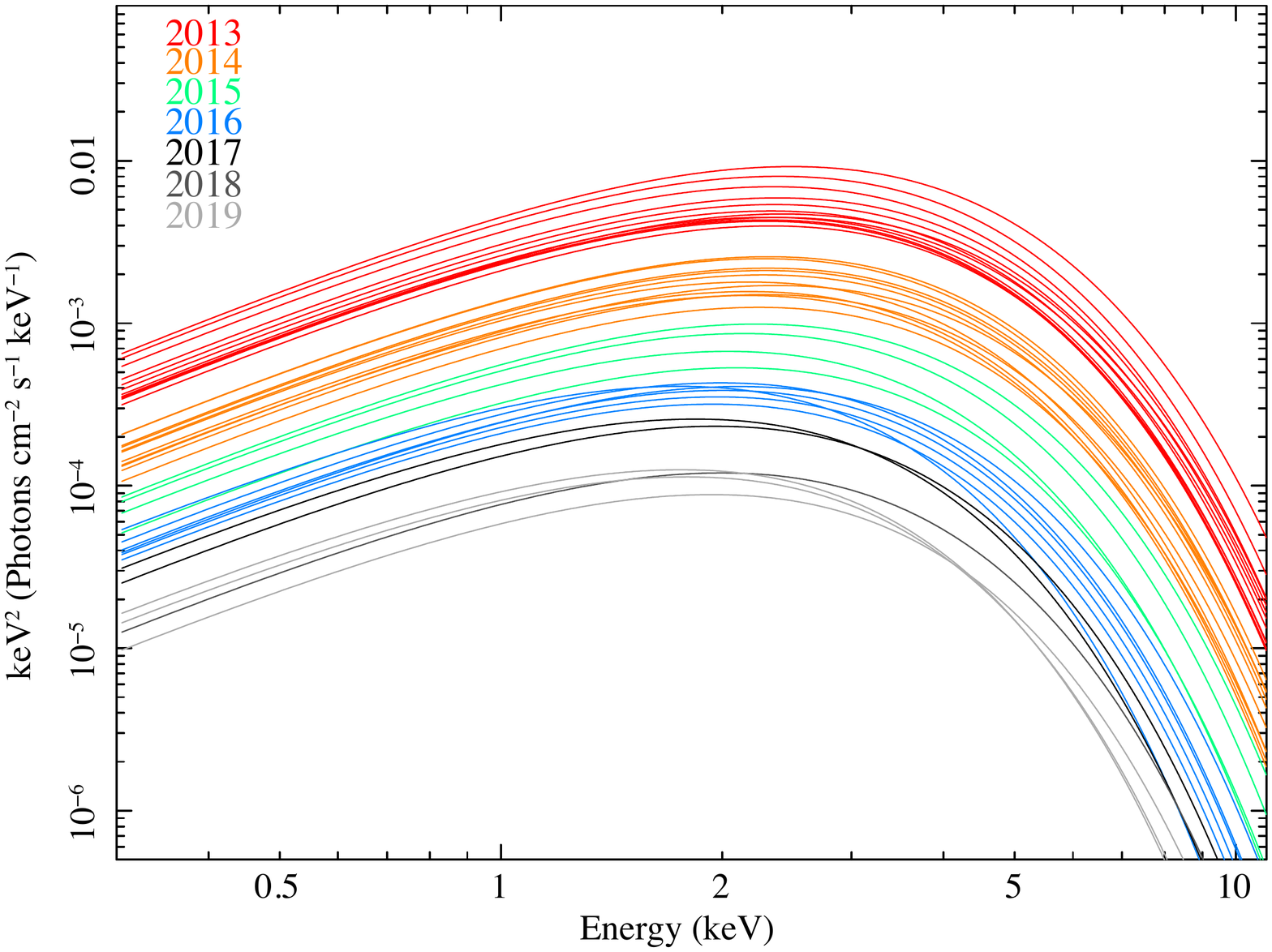}
\caption{The cooling of the surface blackbody of \galmag\, during the outburst decay from April 2013 until August 2019, as observed by \cxo. The left panel displays the spectra fitted as described in \S\ref{subsec:spectra}, while the right panel shows the blackbody models which best fit those spectra.}
\label{fig:cooling}
\end{figure*}

The pulsed radio emission of \galmag\, allowed many interesting measurements, such as the high magnetic field lower limit within the Sgr~A$^{\star}$ environment ($B>8$\,mG; Eatough et al. 2013) and the magnetar proper motion of $236\pm11$\,km\,s$^{-1}$ at a position angle of $22\pm2\deg$ East of North (Bower et al. 2015). 

The closeness of \galmag\ to the Galactic center black hole hampered its observation and monitoring with X-ray instruments with standard few-arcsec spatial accuracy, leaving the study of the long-term behavior of this magnetar to \cxo\, and its superb sub-arcsecond angular resolution. The X-ray monitoring up to 3 years after the outburst activation can be found in several papers (Rea et al. 2013; Kaspi et al. 2014, Coti Zelati et al. 2015, 2017), showing the very slow cooling of this magnetar, as well as large spin down changes during its outburst evolution. 

In this paper, we report on new \cxo\ observations of \galmag\ which complete the characterization of the spectral and timing properties of the Galactic center magnetar until August 2019, covering 6 years of X-ray outburst evolution. In \S~\ref{sec:obs} we describe the observations and the data processing, in \S~\ref{sec:analysis}  we report on the timing and spectral analysis, while the discussion of our results follows in \S~\ref{sec:discussion}.

\begin{figure*}
\begin{center}
\includegraphics[angle=-90,width=12cm]{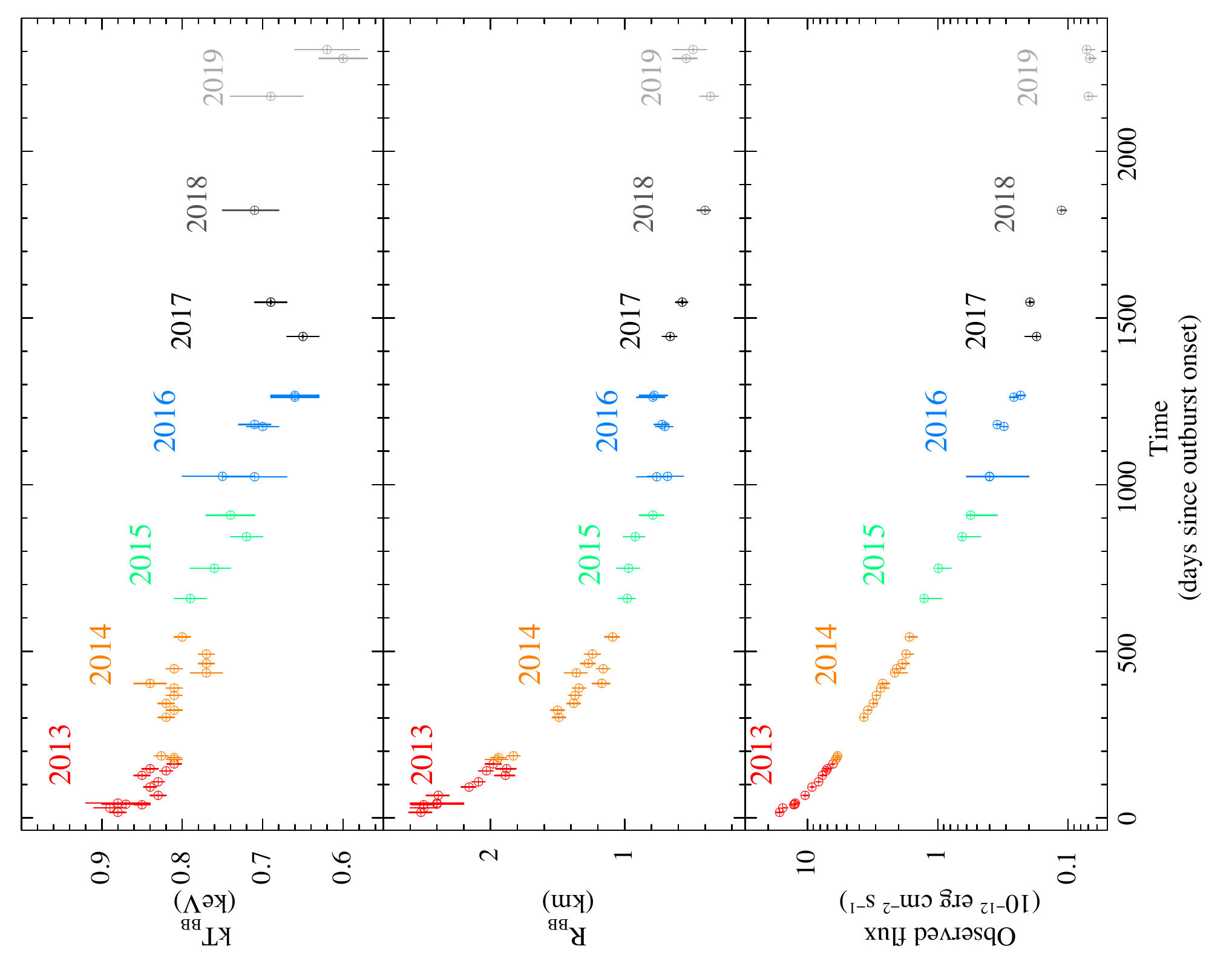}
\caption{Evolution of the spectral parameters of \galmag\, during the outburst decay from 2013 until 2019, as observed by \cxo. Time zero refers to the outburst onset which was on 24/04/2013 (MJD 56406). From top to bottom: the surface temperature, blackbody radius measured at infinity, and the observed 0.3--10\,keV flux.}
\label{fig:param}
\end{center}
\end{figure*}

\section{Observations and data extraction} \label{sec:obs}

The {\em Chandra X-ray Observatory} monitored \galmag\ between 2013 April 29 and 2019 August 19, for a total dead-time corrected on-source exposure time of about 2.3~Ms. Leaving aside the first pointing, which was carried out with the spectroscopic detector of the High Resolution Camera (HRC-S), and 6 pointings with the ACIS imaging array (ACIS-I) that had the source in a very off-axis position, all other observations were performed with the Advanced CCD Imaging Spectrometer spectroscopic array (ACIS-S; Garmire et al. 2003) operated in timed-exposure imaging mode and with faint telemetry format. A few observations were performed using the High Energy Transmission Grating (HETGS), while a 1/8 sub-array was adopted to achieve a time resolution of 0.44104~s, to be sensitive to pulsations at the magnetar spin period of $\sim 3.76$~s. The source was always positioned on the back-illuminated S3 chip. See Table~\ref{tab:log} for more details on the ACIS-S observations used in this work.

All data were processed using the \cxo\ Interactive Analysis of Observations software (\textsc{ciao}, v. 4.11; Fruscione et al. 2006) and the most recent version of the calibration files (\textsc{caldb}, v. 4.8.3). We reduced the data using the same procedures as detailed by Coti Zelati et al. (2017). The source photons were collected from a circular region with a 1.5\arc\ radius, while the background was estimated for each observation against many regions significantly differing in shape, size and proximity to the source (avoiding \sgras, bright transients, and known X-ray sources in the field). A 1.5\arc\ circle at the target position in pre-outburst, archival, ACIS-S observations of the field was also used to assess the correct background level. Pile-up affected the first years of observations (see Coti Zelati et al. 2015 for details), while the latest data sets were not impacted by it, as a result of the decreased source flux. All analyses were restricted to photons having energies between 0.3 and 8 keV. All spectral files, redistribution matrices and ancillary response files were generated via \textsc{specextract}; the spectra were grouped to have at least 50 counts in each energy channel.  All uncertainties in this work are quoted at the 1-$\sigma$ confidence level for a single parameter of interest, unless otherwise specified.

\section{Data analysis}
\label{sec:analysis}

\subsection{Spectral analysis}
\label{subsec:spectra}

We modeled the spectra using \textsc{xspec}\footnote{http://heasarc.gsfc.nasa.gov/xanadu/xspec/} (version 12.10.1f; Arnaud 1996). All spectra were fitted simultaneously with an absorbed blackbody model (see Fig.\,\ref{fig:cooling}), assuming abundances from Wilms, Allen \& McCray (2000) and the photo-electric cross-sections from Verner et al. (1996). In the first 12 spectra (obs IDs from 14702 to 15045, corresponding to the first $\sim 200$ days of the outburst decay) we added a pile-up model (Davis 2001) to account for the spectral distortions induced by pile-up (see Coti Zelati et al. 2015 for more details). 
Given the decay of the source flux, to have enough counts to perform a reasonable spectral analysis with observations where the count rate dropped below 0.001 counts/s, we have merged the observations during each Chandra observing window (time span $<$ 1 month; see also Table~\ref{tab:log}). We checked that no flux or spectral variability were significantly observed among the datasets that were merged.

During the modeling, the hydrogen column density ($\nh$) is tied across all spectra whereas the blackbody temperature and radius are left free to vary. The fit yields a $\chi^2_\nu = 1.01$ for 2718 dof. The inferred column density is $\nh = (1.86\pm0.01) \times 10^{23}$~cm$^{-2}$. We report in Table~\ref{tab:spectralfits}, Figure~\ref{fig:cooling} and \ref{fig:param} the results of the best-fitting black body component, showing the temporal evolution of the spectral parameters and the observed flux. 

The source flux is decreasing very slowly. The blackbody temperature still remains at a relatively high value of $\sim$$0.6$~keV, having cooled down by only 0.3~keV over 6 years of outburst (Figure~\ref{fig:param}). Our fits show that the fading resulted mainly from the shrinking of the black body emitting region, that went from an initial radius of 2.5~km to a small spot of 0.3~km (assuming a distance of 8.3\,kpc).

To have an estimate of the upper limit of the thermal emission from the rest of the neutron star surface, we have added to the model a further black body component fixing its radius to 12\,km (typical value for most neutron star equation of states), and fitting it simultaneously to all observations. We found that the maximum temperature compatible with the observations, for the rest of the surface of \galmag, is $<0.2$\,keV.

\begin{deluxetable*}{ccccccccccc}
\tablecaption{Periods and pulsed fractions measured in the observations
considered here. Upper limits were evaluated at the 3-$\sigma$ confidence level.\label{tab:periods}}
\tabletypesize{\scriptsize}
\tablenum{2}
\tablewidth{0pt}
\tablehead{
\colhead{Obs ID} 	     & \colhead{Start time, t (MJD)} & \colhead{$T_{\rm obs}$ (s)} & \colhead{$N_{\gamma}$}	& \colhead{$\nu_{\rm exp}$ (Hz)} & \colhead{$\sigma$ ($10^{-5}$ Hz)}& \colhead{$P_{\rm max}$} & \colhead{n} & \colhead{$p_{\rm noise}(n)$} & \colhead{$\nu_{\rm meas}$ (Hz)} &  \colhead{Pulsed fraction, $A_{\rm rms}$} 	
	}
\startdata
 20344   &  58228.1571746  & 31859.2 & $136$ & $0.265361$ & $1.9$ & $25.8$ & $5$ & $1.2\times10^{-5}$ & $0.265382(5)$ &    $ 0.45 \pm 0.10$  \\
 20345   &  58230.1728923  & 31148.9 & $103$ & $0.265361$ & $1.9$ & $4.1$  & $5$ & $0.64$             & ...           &    $<0.52$           \\
 20346   &  58232.1643109  & 32686.9 & $115$ & $0.265351$ & $1.9$ & $13.1$ & $5$ & $7.2\times10^{-3}$ & $0.265392(8)$  &     $ 0.32 \pm 0.13 $  \\
 20347   &  58233.1693009  & 35421.5 & $127$ & $0.265360$ & $1.9$ & $5.9$  & $5$ & $0.26$     	      & ...           &    $<0.51$   \\
 21453   &  58571.1871876  & 32341.6 & $76 $ & $0.265279$ & $1.9$ & $4.4$  & $5$ & $0.55$    	      & ...           &    $<0.62$   \\
 21454   &  58572.2500795  & 32903.9 & $93 $ & $0.265279$ & $1.8$ & $4.5$  & $5$ & $0.53$     	      & ...           &    $<0.56$   \\
 21455   &  58573.2396278  & 32275.9 & $83 $ & $0.265279$ & $1.7$ & $12.5$ & $5$ & $9.7\times10^{-3}$ & $0.265296(8)$   &    $ 0.42 \pm 0.14$   \\
 21456   &  58574.2092369  & 32209.6 & $83 $ & $0.265279$ & $2.1$ & $5.9$  & $5$ & $0.26$             & ...           &    $<0.63$   \\
 22230   &  58681.9956763  & 55444.2 & $140$ & $0.265252$ & $2.1$ & $2.3$  & $7$ & $\sim 1$	      & ...           &    $<0.39$   \\  
 20446   &  58685.0180575  & 56953.0 & $142$ & $0.265251$ & $2.2$ & $2.9$  & $7$ & $\sim 1$	      & ...           &    $<0.41$   \\
 20447   &  58690.0866392  & 57053.6 & $153$ & $0.265250$ & $2.2$ & $3.6$  & $7$ & $\sim 1$	      & ...           &    $<0.42$     \\   
 20750   &  58708.9744144  & 25601.0 & $62$  & $0.265252$ & $2.6$ & $5.4$  & $5$ & $0.37$    	      & ...           &    $<0.72$   \\	
 22288   &  58710.9787112  & 25715.7 & $74$  & $0.265251$ & $2.6$ & $4.8$  & $5$ & $0.45$	      &	...           &    $<0.64$   \\ 
 20751   &  58714.9522582  & 25990.8 & $76$  & $0.265250$ & $2.6$ & $5.3$  & $5$ & $0.35$    	      &	...           &    $<0.64$ \\
 \hline
\hline
\enddata
\end{deluxetable*}


\begin{figure*}
\begin{center}
\includegraphics[angle=0,width=9.5cm]{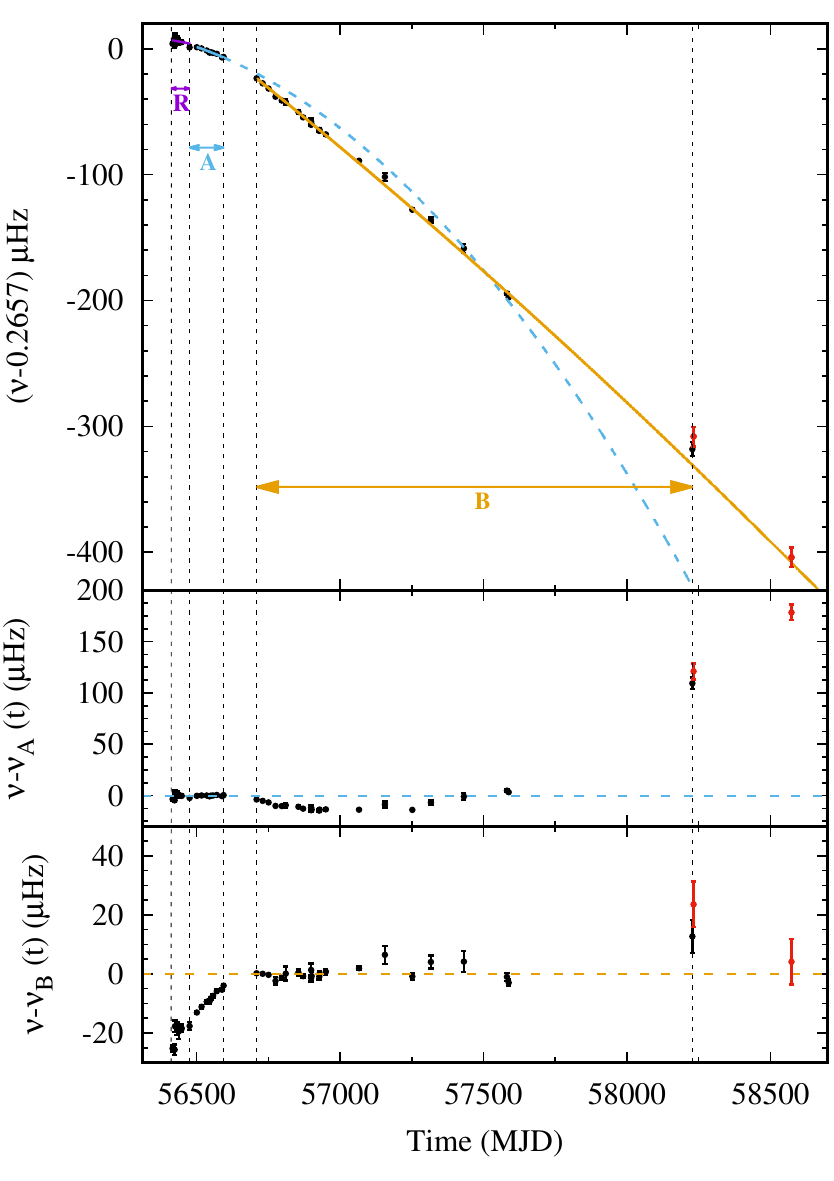}
\caption{Temporal evolution of the spin frequency of \galmag. The magenta solid line shows the phase-connected coherent timing solution given by Rea et al. (2013) valid over the time range labelled R. The  blue solid line is the solution A given by Coti Zelati et al. (2015) valid over the time range labelled A, and the dashed blue line is its extension beyond its range of validity. Residuals with respect to this solution are plotted in the middle panel. The best-fitting model found over the interval MJD~56709.5-58228.2 (labelled B) is plotted with a solid orange line (see also the rightmost column of  Table~\ref{tab:timing}), and residuals with respect to this solution are plotted in the bottom panel. Red points mark the values measured in Obs IDs  20346 and 21455 (not used in the fitted the model). The vertical dashed black lines mark the limits of the ranges of validity of the various solutions. }
\label{fig:timing}
\end{center}
\end{figure*}


\begin{deluxetable*}{lccc}
\tablecaption{Timing solutions. The first solution (labelled R and valid in the range MJD 56411.6-56475.5) is taken from Rea et al. (2013b), the second (labelled A; MJD 56500.1-56594.2) corresponds to Solution A by Coti Zelati et al. (2015) and the third (labelled B; MJD~56709.5-58228.2) is reported in this work and represents
the extension of Solution B given by Coti Zelati et al. (2017) over a longer temporal baseline.
\label{tab:timing}}
\tablenum{3}
\tablewidth{0pt}
\tablehead{
\colhead{Validity Range (MJD)} 	     & \colhead{ R (56411.6--56475.5)} & 	\colhead{ A (56500.1--56594.2)} & \colhead{B (56709.5--58228.2)}}
\startdata
Epoch $T_0$ [ MJD]  	& 56424.5509871 			& 56513.0    	& 56710.0 	\\ \vspace{0.1cm}
$P(T_0)$ [s]      	& 3.7635537(2)				& 3.76363799(7)    	&3.763980(2)     		 		\\ \vspace{0.1cm} 
$\dot{P}(T_0)$ [s s$^{-1}$]  	& $6.61(4)\times10^{-12}$	& $1.360(6)\times10^{-11}$   & $3.02(3)\times10^{-11}$ 	\\ \vspace{0.1cm}
$\ddot{P}$ [s s$^{-2}$] & $4(3)\times10^{-19}$	& $3.7(2)\times10^{-19}$& $0.46(25)\times10^{-19}$ 	 \\ 
\tableline \vspace{0.1cm}
$\nu(T_0)$ [Hz] 	& 0.265706368(14)	&  0.26570037(5)  	 & 0.2656762(2)  	\\ \vspace{0.1cm}
$\dot{\nu}(T_0)$ [Hz s$^{-1}$]  	& $-4.67(3)\times10^{-13}$	& $-9.60(4)\times10^{-13}$   & $-2.13(2)\times10^{-12}$ 	 \\ \vspace{0.1cm}
$\ddot{\nu}$ [Hz s$^{-2}$]  	& $-3(2)\times10^{-20}$	& $-2.6(1)\times10^{-20}$    & $(-0.32\pm0.18)\times10^{-20}$ \\
\tableline  \vspace{0.1cm}
rms residual	&  0.15 s		& 0.396 s	 & $1.9 \times 10^{-5}$ s	\\ \vspace{0.1cm}		
$\chi_\nu^2$ (dof)			& 0.85 (5)		& 6.14 (44)   		 & 2.89 (19)                      			 \\ 
\hline
\hline
\enddata
\end{deluxetable*}

\subsection{Timing analysis}
\label{subsec:timing}

To derive updated ephemeris, we searched the new data for coherent pulsations in the range of frequencies expected according to the timing solution  given by Coti Zelati et al. (2017; see the rightmost column of Table 2 in that paper, which represents the updated version of the solution labelled as B by Rea et al. 2013 and Coti Zelati et al. 2015).
 We evaluated  a Fourier power density spectrum for each of the observations listed in Table~\ref{tab:periods}, and 
restricted the search for a pulsed signal in the range of the pulsar spin  frequencies expected according to the solution, $\nu_{\rm exp}(t)=\nu_{\rm CZ}(T_0)+\dot{\nu}_{\rm CZ}(T_0) (t-T_0)+1/2 \ddot{\nu}_{\rm CZ} (t-T_0)^2$. Here, $\nu_{\rm CZ}(T_0)$, $\dot{\nu}_{\rm CZ}(T_0)$ and $ \ddot{\nu}_{\rm CZ} $ are the ephemeris measured by Coti Zelati et al. (2017) over the time interval 56709.5--57588.5 MJD, $T_0=56710.0$~MJD is the reference epoch of that timing solution, and $t$ is the start time of the actual observation considered (see Table~\ref{tab:periods}). The frequencies considered in the search ranged from $\nu_{\rm exp}(t) - 3 \sigma$ to $\nu_{\rm exp}(t) + 3 \sigma$. Here $\sigma=[\sigma(\nu_{\rm exp}(t))^2+\sigma(\nu_{IFS})^2]^{1/2}$ is the quadratic sum of the uncertainty on the expected frequency  $\sigma(\nu_{\rm exp}(t))$, derived by propagating the uncertainties of the older timing solution measured by Coti Zelati et al. (2017) to the epoch of the observations considered here, and the uncertainty on the frequency measured in the new observations $\sigma(\nu_{IFS})=1/(2T_{\rm obs})$, equal to half the spacing between independent Fourier frequencies (see Table~\ref{tab:periods}). We evaluated the number of trials needed to sample the expected range of frequencies as $n=1+3\sigma/\sigma(\nu_{IFS})$, and obtained values ranging from $n=5$ to $7$. This is equivalent to performing a search for periodicities with a flat prior on frequencies in the $\pm3\sigma$ interval determined from the previous timing solution.

\begin{figure}
\begin{center}
\includegraphics[angle=0,width=8cm]{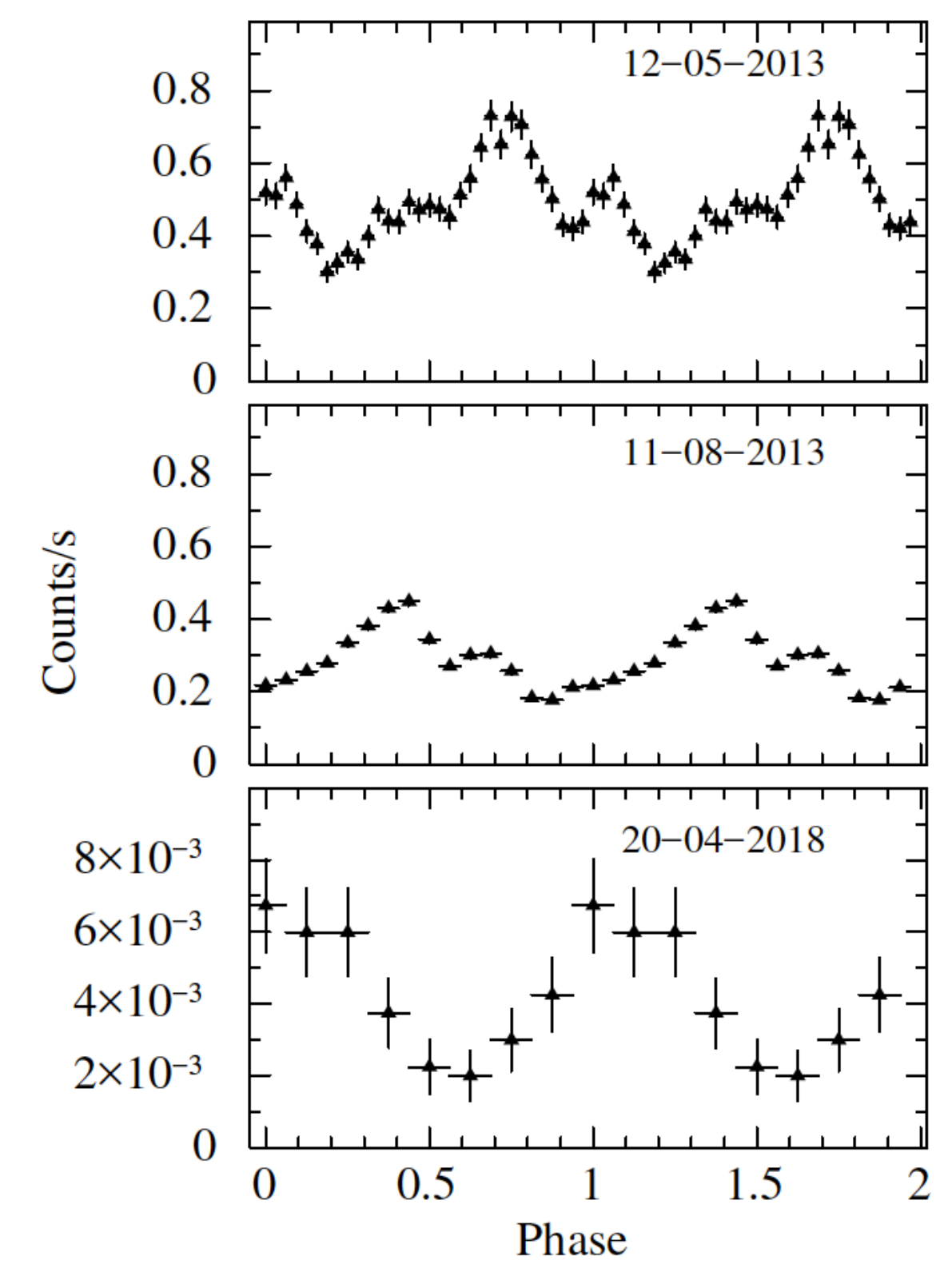}
\caption{Pulse profiles for three \cxo\, observations (IDs: 14702, 15042, and 20344; see Table\,\ref{tab:log}), each one in a different ephemerides validity range (see Table\,\ref{tab:timing}). }
\label{fig:profiles}
\end{center}
\end{figure}

The maximum Fourier power density observed in the considered range, $P_{\rm max}$, and the probability $p_{\rm noise}(n)$ of being due to white noise weighted for the number of trials $n$, are given in Table~\ref{tab:periods}. We chose to consider that a detection is statistically significant if  $p_{\rm noise}(n)$ is lower than $2.7\times10^{-3}$. Only during observation ID 20344 the pulsed signal had a very low probability of being due to noise, $1.2\times10^{-5}$, which is equal to cumulative probability beyond $4.4\sigma$ of a standard normal distribution. During observations IDs 20346 and 21455, a signal with a low probability of being due to noise, but still larger than the  3-$\sigma$ white noise threshold, was found. We determined the frequency $\nu_{\rm max}$ in each of these observations by performing an epoch folding search sampling the pulse in 8 bins, and fitting the peak of the pulse variance with a Gaussian distribution. The uncertainties on the frequency $\nu_{\rm meas}$ listed in Table~\ref{tab:periods} were obtained following Leahy et al. (1987). We evaluated the rms pulsed fraction $A_{\rm rms}$ fitting the pulse profile obtained folding the time series at $\nu_{\rm max}$ with sinusoid The bottom panel of Fig.~\ref{fig:profiles} shows the pulse profile observed during observation ID 20344. For the remaining observations in which no signal was significantly detected, we evaluated upper limits on the pulsed fraction given in Table~\ref{tab:periods} at the 3-$\sigma$ confidence level (see, e.g., Vaughan et al. 1994). 

\begin{figure}
\begin{center}
\includegraphics[angle=0,width=8.5cm]{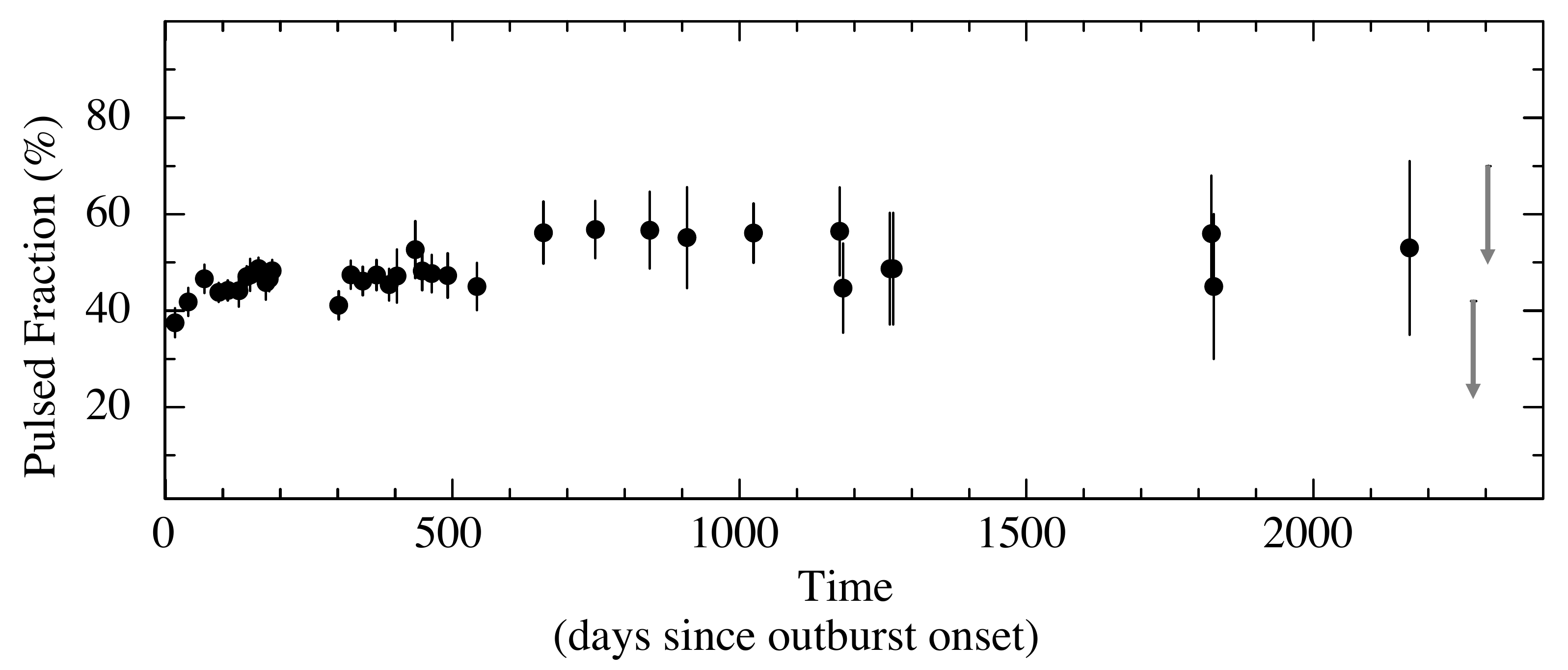}
\caption{Pulsed fraction evolution as a function of time. We report the upper limits as derived in Tab.~\ref{tab:periods} for the last six observations, and averaging over close observations.}
\label{fig:pf}
\end{center}
\end{figure}

The only high significance measurement of the pulsar spin frequency (Obs. ID 20344) is fitted together with the periods determined by Coti Zelati et al. (2017) in the interval MJD~56709.5--57588.5 with a quadratic function $\nu(t)=\nu(T_0)+\dot{\nu}(T_0) (t-T_0)+1/2 \ddot{\nu} (t-T_0)^2$. The derived (non phase-connected) solution is reported in the rightmost column of Table~\ref{tab:timing}. The top panel of Figure~\ref{fig:timing} shows the frequency evolution over the entire dataset available. The phase-connected solution given by Rea et al. (2013; valid over the interval MJD~56411.6-56475.5, labelled R) and the solution found by Coti Zelati et al. (2015; valid over the interval MJD~56500.1-56594.2, labelled A) are plotted as a magenta and a blue solid line, respectively. The timing solution derived in this work is plotted as an orange solid line, and is valid across the interval marked by the horizontal arrow labelled as B. The two lower significance frequency measurements ( 20346 and 21455) were also plotted (although not fitted) with red symbols to show that they also follow the best-fitting trend. Residuals with respect to solution A and B are given in the middle and bottom panel, respectively. The addition of the data presented here confirms that the solution A is unable to model the evolution of the frequency measured after MJD~56709. Figure~\ref{fig:pf} shows the evolution of the pulsed fraction in time, that settled at a value of $\sim$50\%.


\section{Discussion}
\label{sec:discussion}
  
After more than a decade of systematic monitoring of magnetar outbursts, we can summarize the main common features we have observed so far: (i) a sudden increase and a relatively fast decrease (first year maximum) of the X-ray flux and inferred surface black body temperature; (ii) a subsequent gradual shrinking of the inferred size of the thermally emitting region and a gradual decrease of its temperature; (iii) a fast softening (and often quick disappearance) of a non-thermal X-ray component, sometimes reaching a few hundreds keV; (iv) certain variability of the spin down rate during the outburst decay.

However, having now monitored a few magnetar outbursts for several years, the overall picture of the late-time outburst evolution displays a clear diversity after the first year or two. After the initial period of fast cooling, sources appear to behave in different ways, as revealed by the more recent observations. At the beginning of the outburst, the source evolution is compatible with the two most common theoretical scenarios: cooling of extra heat deposited internally, or imprints of magnetospheric currents. In both these scenarios, the physical trigger is arguably the internal failure of the crust due to excessive magnetic stresses: cracking (Thompson \& Duncan 1995; Perna \& Pons 2011) or thermoplastic waves propagation (Beloborodov \& Levin 2014). The two scenarios have different phenomenological implications. In particular, if a high temperature ($\gtrsim$ 0.5 keV) is maintained for long time ($\gtrsim$ 1 year), like in the \galmag, the heat diffusion timescales make the ``cooling from inside" scenario unfeasible, unless a continuous, shallow deposition of heat happens. On the other hand, this continuous deposition of heat would be difficult to justify and energetically unrealistic. Instead, in the twisted bundle scenario, coronal loops persist over timescales which can vary from many months to decades (Beloborodov 2009). Currents circulate in the interior and exterior of the star, and heat the surface in a hotspot by Joule dissipation in the external layers. As currents get dissipated, the hotspot shrinks and the luminosity decreases.

However, the case of the X-ray outburst of \galmag\, is not easily ascribable to any of the above-mentioned scenarios. On the one hand, the rather high long-lasting temperature of the emitting hot spot ($kT_{{\rm BB}}\sim 0.9-0.6$\,keV) and its gradual shrinking ($R_{{\rm BB}}\sim2.5-0.3$\,km) are incompatible with the internal cooling scenario. On the other hand, the absence of a non-thermal component over most of the outburst (it was observed only for a few months after the trigger; Kaspi et al. 2014) does not easily support the long-term presence of a powerful magnetic bundle heating the surface from outside (unless arguments related to an unfavorable beaming of the up-scattered photons are invoked).

It is interesting to note that studying  the implications of the crust-magnetosphere coupling, Akgun et al (2018) found that allowing strong currents passing through the last hundred meters of the
surface (the envelope) where the magnetic diffusivity is high, results in a considerable amount of energy being deposited very close to the stellar surface, where Joule heating is more efficient (as opposed to
the interior, where neutrino losses are significant). They show that under certain circumstances the effective surface temperature could increase locally from 0.1 keV to 0.6 keV. Therefore, more attention must be paid to understand how long-lived magnetospheric currents close the circuit through the star, which may be in the future the key to understand the long-lived high temperature of \galmag.

The spin period derivative of \galmag\ has increased overall by a factor of $\sim$\,4 along the outburst decay. The pulsed fraction has increased only slightly during the first year of the outburst, and maintained a rather constant value within the 40-50\% range over the subsequent $\sim4$ years, until the epochs of the most recent detections of pulsed X-ray emission.

Time variability in the spin-down rate is an ubiquitous property for magnetars in outburst (Kaspi \& Beloborodov 2017). An accurate assessment on the torque evolution in these sources is often hampered by the sparse observational coverage along their outburst decay. Nevertheless, there is no evidence for a common trend among the magnetar sample. Sources such as 1E 1048.1-5937 and the radio magnetars \aa, \psr\ and \xte, to mention a few, showed unique dramatic changes, and were also observed to undergo glitch and anti-glitch events (Archibald et al. 2020 and references therein).
Variations in the spin-down rate of magnetars are believed to be driven by the evolution of the magnetic bundle in the magnetosphere. The basic picture predicts that the spin-down torque should initially increase as the twist grows, then decrease and eventually recover the pre-outburst value as the bundle dissipates (Beloborodov 2009). However, this scenario can account for the observed torque evolution only in a few cases (see, e.g., Pintore et al. 2016), while it does not provide a straightforward explanation for the extremely varied phenomenology of most magnetars. As a matter of fact, detailed simulations would be needed to investigate how the evolving magnetic twist determines the torques in most magnetars along their outburst.

\begin{figure}
\begin{center}
\includegraphics[angle=0,width=1\columnwidth]{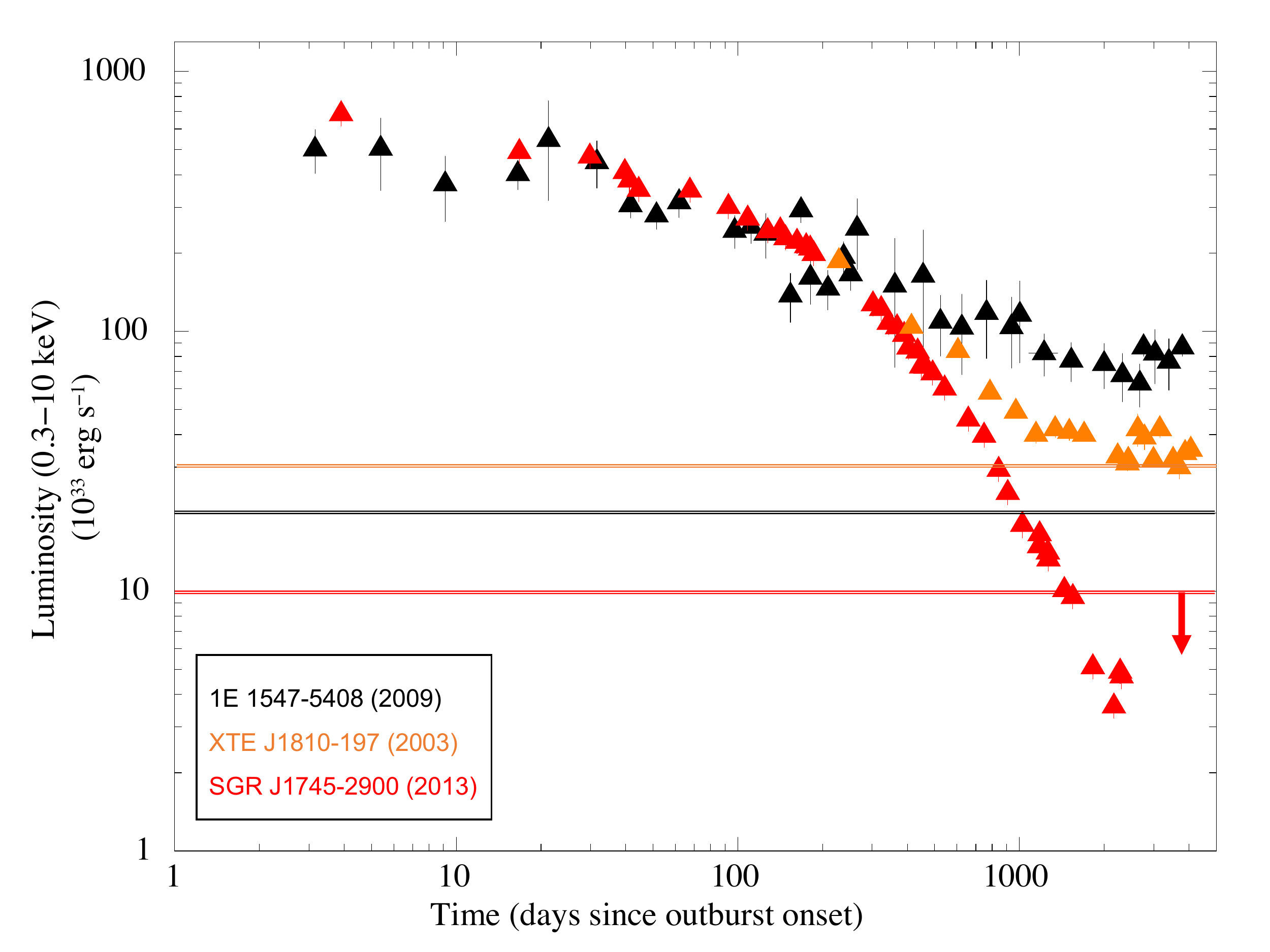}
\caption{Comparison between the outburst luminosity evolution of all radio magnetars having yearly-long coverage of their outbursts. The black and orange horizontal lines refer to the quiescent luminosity values observed for \aa\ and \xte. The red horizontal line is the pre-outburst luminosity upper limit for \galmag\, as derived with archival \cxo\, observations (see text for details). Time zero refers to the outburst onset which was 24/04/2013, 01/01/2003, and 22/01/2009, for \galmag, \xte, and \aa, respectively. }
\label{fig:comparison_lum}
\end{center}
\end{figure}

\subsection{Comparison between radio emitting magnetars}

A further element of complexity comes from the comparison of detected radio emission in coincidence with magnetar outbursts. They are normally absent, however in some peculiar cases they appear after the outburst X-ray peak, while in other cases regular radio pulses are present and get quenched during X-ray bursts (Archibald et al. 2017). This erratic behavior is currently difficult to explain, given also our poor knowledge of the physical mechanism at the base of pulsars' radio emission. It could be related, for instance, to changes in the rotational energy losses, to the (de)-activation of physical conditions allowing the emission and propagation of radio waves, or to a change of the beaming direction due to magnetic reconfiguration.

Four magnetars are known to emit radio pulsations, to which we should add the radio pulsar, \hbpsrtwo\, which recently showed magnetar-like activity (Archibald et al. 2018). The first radio magnetar discovered was \xte, which remained radio active for almost 5 years after the onset of its X-ray outburst (see Fig.~\ref{fig:comparison_lum} for its late-time X-ray decay). After that period, it became undetectable in the radio during the rest of the X-ray decay which lasted about 9 years (Camilo et al. 2016). 
After a period of quiescence (Pintore et al. 2019), the source underwent a new powerful X-ray outburst at the end of 2018, and radio pulsations were again detected (e.g. Gotthelf et al. 2019).
Another radio magnetar is \psr, the only magnetar discovered at radio wavelengths without prior knowledge of an X-ray counterpart (Levin et al. 2010). At the time of discovery, its X-ray flux was decaying from an outburst possibly started around 2007, and no X-ray pulsations could be detected (Anderson et al. 2012). Detectable radio emission ceased in 2014 and, despite frequent monitoring, the pulsar remained undetectable until late 2016 (Scholz et al. 2017), when it underwent a second outburst in the X-ray and radio with detectable pulsations in both bands (Camilo et al. 2018; its X-ray evolution is not reported in Fig.~\ref{fig:comparison_lum} because only the first year of X-ray data are currently available). 

The other radio magnetars are \aa\, and \galmag, that have remained very active both in radio and X-rays for several years after the outburst onset. \aa\, is still extremely luminous in the X-ray band with respect to its pre-outburst quiescent level (Coti Zelati et al. 2020). Its long-term radiative properties are challenging the internal cooling model, but are still compatible with the magnetic bundle model given the strong non-thermal component that is still present in its spectrum 10 years after the outburst onset (see Fig.~\ref{fig:comparison_lum}). 

It is interesting to compare the long term X-ray outburst of these radio-magnetars: all follow a very slow cooling, taking several years, but with very different decays.
The level of the X-ray peaks of \aa\, and \galmag\, is similar, with a clear flux decrease during the first year. However, \galmag~ keeps on fading, being now over two orders of magnitude fainter than at the beginning of the outburst, and having now reached a fainter level of quiescence than the limits we could derived from the pre-outburst \cxo\, observations (red line and arrow in Fig.~\ref{fig:comparison_lum}). On the other hand, \aa\, has maintained a high level of flux and inferred temperature (Coti Zelati et al. 2020), compatible with being constant in the past few years. Note that if we had lacked deep limits on the pre-outburst luminosity (see Fig.~\ref{fig:comparison_lum}), we would have defined the current state as the \aa's standard quiescent luminosity. This is not the case for \galmag\, which however shows the same uncommon length of fading timescales. Concerning \xte, the early times of its outburst in 2003 were missed (and only observed with RXTE above 2\,keV until about 200 days after the estimated onset epoch). Furthermore, its new outburst started at the end of 2018 is still too recent to be compared with the other events presented here. 

\section{Conclusions}

The Galactic center magnetar, \galmag, keeps fading with a relatively slow but steady rate. The high temperature of the emitting region of \galmag\, after 6 years of outburst decay, and the shrinking of its emitting radius over the outburst evolution disfavor the internal cooling scenario for this event. On the other hand, the purely thermal emission, with no sign of a non-thermal component over the past 5 years, disfavor a long-lived magnetospheric bundle as the source of heat powering the emission of this peculiar outburst, which clearly represents a very peculiar event in the magnetar outburst population.

\acknowledgments

 NR, DV and AB are supported by the H2020 ERC Consolidator Grant "MAGNESIA" under grant agreement Nr. 817661 (PI: Rea). NR, FCZ, DV, AB and DFT also acknowledge support from grants SGR2017-1383 and PGC2018-095512-BI00. FCZ is supported by a Juan de la Cierva fellowship. AP acknowledges financial support from the Italian Space Agency and National Institute for Astrophysics, ASI/INAF, under agreements ASI-INAF I/037/12/0 and ASI-INAF 2017-14-H.0. D.H. acknowledges support from the Natural Sciences and Engineering Research Council of Canada (NSERC) Discovery Grant, the Fonds de recherche du Qu\'ebec–Nature et Technologies (FRQNT) Nouveaux Chercheurs program, and the Canadian Institute for Advanced Research (CIFAR). JAP acknowledges support by the Generalitat Valenciana (PROMETEO/2019/071) and by Agencia Estatal de Investigaci\'on (PGC2018-095984-B-I00). GP is supported by the H2020 ERC Consolidator Grant “Hot Milk” under grant agreement Nr. 865637. LS acknowledges financial contributions from ASI-INAF agreements 2017-14-H.O and  I/037/12/0 and from “iPeska” research grant (P.I. Andrea Possenti) funded under the INAF call PRIN-SKA/CTA (resolution 70/2016). We acknowledge support from the PHAROS COST Action (CA16214). This article is based on data obtained with the Chandra X-ray Observatory, and on softwares and tools provided by the High Energy Astrophysics Science Archive Research Center (HEASARC), which is a service of the Astrophysics Science Division at NASA/GSFC and the High Energy Astrophysics Division of the Smithsonian Astrophysical Observatory. We are grateful to Giovanni Fazio, Joseph Hora, Gordon Garmire and Steven Willner, and their proposal co-Is, for sharing their data. 

\bibliographystyle{apj}

\bibliographystyle{aasjournal}



\end{document}